\documentclass[%
 reprint,
nobibnotes,
amsmath,amssymb,
aps,
pra,
]{revtex4-1}

\usepackage{graphicx}
\usepackage{dcolumn}
\usepackage{bm}
\usepackage{braket}
\usepackage{natbib}
\usepackage{xcolor}
\usepackage{lipsum}

\begin{document}


\title{Quantum jump simulation in three-level systems using photonic Gaussian modes}

\author{A. C. Cardoso\textsuperscript{1}}
\author{J. G. L. Cond\'e\textsuperscript{1}}
\author{B. Marques\textsuperscript{2}}
\author{J. S. Cabral\textsuperscript{3}}
\author{S. P\'adua\textsuperscript{1}}

\affiliation{\textsuperscript{1}Departamento de F\'isica, Universidade Federal de Minas Gerais, Belo Horizonte, MG 31270-901, Minas Gerais, Brazil \\ \textsuperscript{2}Centro de Ci\^encias Naturais e Humanas, Universidade Federal do ABC, Santo André, Brazil  \\ \textsuperscript{3}Instituto de F\'isica, Universidade Federal de Uberl\^andia, 38400-902, Uberl\^andia, MG, Brazil
}

\date{\today}

\begin{abstract}
Multi-level quantum systems loose coherence due to quantum jumps or spontaneous decay between their internal levels. Here we propose a way to simulate experimentally a three-level system under quantum jump using a three-mode photonic system. We simulated three different dynamics of spontaneous decay in a three-level atomic system: cascade decay, $\Lambda$ decay and $V$ decay. With an attenuated light coherent source at the photon level, we prepared a photonic qutrit state encoded in the parallel path of Gaussian modes. By exploring periodical phase modulation in spatial light modulator, the corresponding dynamical maps for quantum jumps were implemented in terms of the Kraus operator decomposition. With image measurements with an intensified charged-coupled device (ICCD) camera we obtain the diagonal elements of the initial state density matrix. Measuring the image and interference patterns of the evolved qutrit state we verified experimentally the variation of the populations and the decoherence effects caused by the quantum jumps in the three-level system for the three-level decay configurations.

\end{abstract}

\pacs{Valid PACS appear here}
\maketitle

\section{\label{sec:level1}Introduction}

Over the last decades, quantum optical systems have proved to be an useful option for fundamental tests of quantum mechanics and implementation of quantum information and quantum computation protocols \cite{flamini, slussarenko,hensen,ralph}.
Quantum systems are in general not isolated systems and are most of the time subjected to uncontrolled interactions to an external quantum system and/or to its surrounds (environment) \cite{petruccione,alicki,carmichael}. The system interactions with the environment cannot be described by unitary operations acting on the system and lead to dissipation (loss of energy to 
the environment) and decoherence (loss of state coherence) \cite{REVZurek, marquardt,sc}, consequently to a degradation in quantum protocols  \cite{SCIREPMarques,SCIENCEAlmeida,SCIFarias,unden}. Different uncontrolled interactions, referred here as noise, that affect the coherence of the quantum system state or entanglement in bipartite or multipartite state systems have been simulated quantically, more specifically, dephasing, amplitude damping and Pauli noise \cite{SCIREPMarques,mataloni,aolita,almeida,laurat,barbosa,xu,blatt}.
Another important source of noise is the fundamental process called quantum jump, where a quantum system evolves stochastically in an abrupt unpredictable operation. Quantum jump is an essential topic in the interpretation of quantum dynamics \cite{cook} and has been part of the historical debates about the quantum mechanics fundamentals \cite{dick}. This stochastic process, first proposed by Niels Bohr \cite{bohr}, has been observed in a single ion \cite{sauter},  molecule \cite{basch}, electron in a trap \cite{peil}, photon in a cavity \cite{gleyzes}, and in artificial atoms \cite{solid}. Present in the photoelectric effect \cite{photo} and in the spontaneous atom decay this process has an important role in laser cooling \cite{lasercooling}. Although in free atoms the jump occurs on the lifetime of the excited state this time can be increased or shortened by surrounding the atoms with a cavity \cite{enhance}. More recently, quantum jump was tracked in time by following the population of an auxiliary level coupled to the ground state of an artificial atom \cite{minev}.

The transverse profile of optical beams at photon count scenario has been a rich platform for preparing discrete quantum states and for investigating quantum information theories and protocols \cite{sonja}. Photon beams have been prepared in high dimension entangled states of their optical angular momenta in Hermite and Laguerre-Gauss modes \cite{padgett, langford, zeilinger}. Some useful optical systems make use of the photon transverse momentum, which can be discretized by slits \cite{PRLNeves,lima09,stevereview} or in different photon paths  with the aid of interferometers \cite{PRABorges,guo} for preparing one-, two- or four-photon quantum states in slits modes or Gaussian modes \cite{leo1,PRLNeves,paula,holanda,pierre}. The use of the spatial light modulator (SLM) in these optical systems allows the photon state to be manipulated in different ways and can be used to implement a wide range of quantum operations \cite{padgett, lima09,PRAMarques,PRABorges,miguelleo,boyd}.

One crucial advantage of these physical quantum systems is that they are able to simulate much more complex quantum systems \cite{nori,blatt2}. Several experiments explore this fact to study different kinds of quantum system dynamics \cite{SCIREPMarques,SCIENCEAlmeida}. These articles aim to simulate experimentally an decaying dynamic of a three-level system by preparing a photonic qutrit path state and letting the photon beam in the qutrit state be modified by periodical phase modulation produced by a SLM. The proposal is to implement the spontaneous decay dynamics of a three-level atomic system in different configurations: cascade decay, $\Lambda$ decay and $V$ decay. 

An excited atomic system may undergo spontaneous decay through the interaction with the vacuum state of the electromagnetic field. This kind of system typically experiences a time-dependent exponential decay, in which the probability of an excited state $\ket{i}$ decaying to $\ket{j}$ is given by $p_{ij}=1-e^{-\gamma_{ij}t}$, where $\gamma_{ij}$ is the spontaneous decay rate between levels $i$ and $j$. This process causes a reduction in the population of the excited state and also decoherence. The spontaneous decay of the state may spoil the implementation of quantum information protocols in atomic systems \cite{PRLLukin}. 
This papers is organized as follows: in section~\ref{sec: Kraus} we detail the spontaneous dynamic for three-level systems; in section~\ref{sec:Simulation} we show the experimental setup used to encode and simulate the decoherence; in section~\ref{sec:level3} is presented and discussed the results and in section \ref{conclusions} we conclude and discussed perspectives.
\section{\label{sec: Kraus}Three-level decay dynamics}
\begin{figure}[h]
        \includegraphics[scale=0.15]{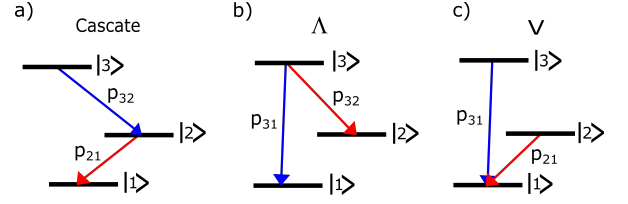}
        \caption{Different configurations of spontaneous decay dynamic in a three-level system: cascade decay, $\Lambda$ decay and $V$ decay.}
        \label{fig:levelconfig}
    \end{figure}

In a three-level system, we consider the three configurations of spontaneous decay dynamics: cascade decay, $\Lambda$ decay and $V$ decay, each one with a forbidden transition and without degeneracy, which are represented in the Fig. \ref{fig:levelconfig}.

In the theory of quantum open system, for all these decay dynamics, the time evolution of the three-level atomic system is obtained by the application of a dynamical map. In this case, we will approach the dynamical map in terms of their Kraus decompositions. The density operator evolved in time, $\rho(t)$, can be written as $\rho(t)=\sum_{i}K_i\rho_0K^{\dagger}_i$, where $K_i$ are the Kraus operators calculated for each decay type, $ \sum_{i}^{} K_i K^{\dagger}_i = I $ ($I$ is the identity operator) and $\rho_0=\ket{\psi_0}\bra{\psi_0}$ is the initial density operator ($\ket{\psi_0}=\ket{\psi\left(t=0\right)}$). The Kraus operators for each configuration are presented in Table \ref{tab:Kraus_operators}.

The density operator evolved in time,  $\rho(t)$, can be represented by a density matrix in terms of the level states $\ket{i}$ ($i = 1, 2, 3$). We obtain the density matrix evolved in time, Eqs. (\ref{eq:densmatrix1}-\ref{eq:densmatrix3}), where $\rho(t)=\sum_{ij}\rho_{ij}\ket{i}\bra{j}$ and $\rho_{ij}$ are the matrix elements:
\begin{widetext}
     \begin{eqnarray}
     \label{eq:densmatrix1}
     \rho(t)&=&{M}_{C}(\rho_0)=\frac{1}{I_T}
     \begin{pmatrix}
        I_1+I_2p_{21}+I_3p_{32}p_{21}   &   \sqrt{I_1I_2}\sqrt{1-p_{21}}   &   \sqrt{I_1I_3}\sqrt{1-p_{32}}\\
        \sqrt{I_1I_2}\sqrt{1-p_{21}}   &   I_2(1-p_{21})+I_3p_{32}(1-p_{21})   &  \sqrt{I_2I_3}\sqrt{1-p_{21}}\sqrt{1-p_{32}}\\
        \sqrt{I_1I_3}\sqrt{1-p_{32}} & \sqrt{I_2I_3}\sqrt{1-p_{21}}\sqrt{1-p_{32}} &   I_3(1-p_{32})
    \end{pmatrix},\\ 
    \label{eq:densmatrix2}
    \rho(t)&=&{M}_{\Lambda}(\rho_0)=\frac{1}{I_T}
     \begin{pmatrix}
        I_1+I_3p_{31}   &   \sqrt{I_1I_2}   &   \sqrt{I_1I_3}\sqrt{1-p_{31}-p_{32}}\\
        \sqrt{I_1I_2}   &   I_1+I_3p_{32}   &  \sqrt{I_2I_3}\sqrt{1-p_{31}-p_{32}}\\
        \sqrt{I_1I_3}\sqrt{1-p_{31}-p_{32}} & \sqrt{I_2I_3}\sqrt{1-p_{31}-p_{32}} &   I_3(1-p_{32}-p_{31}
    \end{pmatrix},\\
    \label{eq:densmatrix3}
    \rho(t)&=&{M}_{V}(\rho_0)=\frac{1}{I_T}
     \begin{pmatrix}
         I_1+I_2p_{21}+I_3p_{31}   &   \sqrt{I_1I_2}\sqrt{1-p_{21}}   &   \sqrt{I_1I_3}\sqrt{1-p_{31}}\\
        \sqrt{I_1I_2}\sqrt{1-p_{21}}   &   I_2(1-p_{21})   &  \sqrt{I_2I_3}\sqrt{1-p_{21}}\sqrt{1-p_{31}}\\
        \sqrt{I_1I_3}\sqrt{1-p_{31}} & \sqrt{I_2I_3}\sqrt{1-p_{21}}\sqrt{1-p_{31}} &   I_3(1-p_{31})
    \end{pmatrix},
    \end{eqnarray}
\end{widetext}
where we assume the initial state to be pure and equal to $\ket{\psi_0}$ that can be rewritten as~$\ket{\psi_0}=\frac{1}{\sqrt{I_T}}\left(\sqrt{I_1}\ket{1}+\sqrt{I_2}\ket{2}+\sqrt{I_3}\ket{3}\right)$, $I_T = \sum^{3}_{\ell=1} I_\ell$ and $M_k$ is the map for the $k$ configuration  ($k = C,~\Lambda,~V$).

 \begin{table*}[t!]
		\renewcommand{\arraystretch}{1.5} 
		\caption{Kraus operators for cascade decay, $\Lambda$ decay and $V$ decay.} \label{tab:Kraus_operators}
		\begin{center}
		\begin{tabular}{c c c c}
		    \hline
			\hline
			 & Cascade	& $\Lambda$ & $V$  \\
			 \hline
			$K_0$	& \begin{math} \begin{pmatrix} 1 & 0 & 0 \\ 0 & \sqrt{1-p_{21}} & 0 \\ 0 & 0 & \sqrt{1-p_{32}}  \end{pmatrix}\end{math} & %
			\begin{math}\begin{pmatrix}  1 & 0 & 0 \\ 0 & 1 & 0 \\ 0 & 0 & \sqrt{1-p_{32}-p_{31}}  \end{pmatrix}\end{math} & %
			\begin{math}\begin{pmatrix}  1 & 0 & 0 \\ 0 & \sqrt{1-p_{21}} & 0 \\ 0 & 0 & \sqrt{1-p_{31}}  \end{pmatrix}\end{math} \\ %
			$K_1$	& \begin{math}\begin{pmatrix}  0 & \sqrt{p_{21}} & 0 \\ 0 & 0 & 0 \\ 0 & 0 & 0  \end{pmatrix}\end{math} & %
			\begin{math}\begin{pmatrix}  0 & 0 & 0 \\ 0 & 0 & \sqrt{p_{32}} \\ 0 & 0 & 0  \end{pmatrix}\end{math} & %
			\begin{math}\begin{pmatrix}  0 & \sqrt{p_{21}} & 0 \\ 0 & 0 & 0 \\ 0 & 0 & 0  \end{pmatrix}\end{math} \\ %
			$K_2$	& \begin{math}\begin{pmatrix}  0 & 0 & 0 \\ 0 & 0 & \sqrt{p_{32}(1-p_{21})} \\ 0 & 0 & 0  \end{pmatrix}\end{math} & %
			\begin{math}\begin{pmatrix}  0 & 0 & \sqrt{p_{31}} \\ 0 & 0 & 0 \\ 0 & 0 & 0  \end{pmatrix}\end{math} & %
			\begin{math}\begin{pmatrix}  0 & 0 & \sqrt{p_{31}} \\ 0 & 0 & 0 \\ 0 & 0 & 0  \end{pmatrix}\end{math} \\ %
			$K_3$	& \begin{math}\begin{pmatrix}  0 & 0 & \sqrt{p_{32}p_{21}} \\ 0 & 0 & 0 \\ 0 & 0 & 0  \end{pmatrix}\end{math}  &  & \\
			\hline
			\hline
		\end{tabular}
		\end{center}
    \end{table*}
\section{\label{sec:Simulation}The simulation of the dynamics}
\subsection{\label{subsec:Exp_setup} Qutrit state preparation and operation setup}

\begin{figure*}
  \centering
   \includegraphics[scale=0.5]{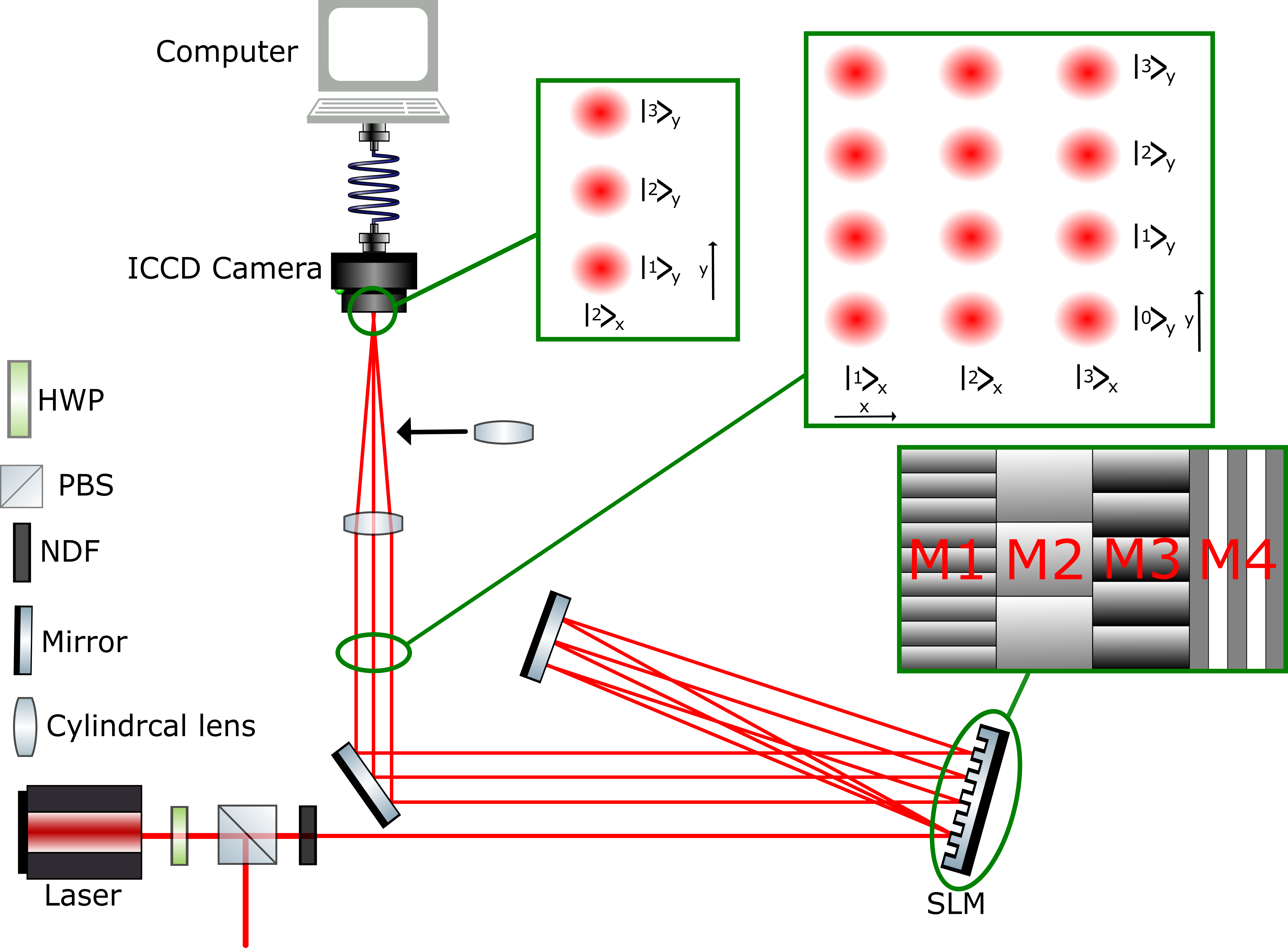}
     \caption{Experimental setup used to prepare and to implement the operations on a qutrit parallel Gaussian beam state. A laser beam passes through a half wave-plate (HWP), a polarizer beam splitter (PBS) and a neutral density filter (NDF). This is necessary for filtering the polarization state and to attenuate it to the single photon regime. The laser beam reaches a SLM which the screen is divided in four regions, each one with a periodical phase diffraction grating. The attenuated laser beam is initially diffracted  by a periodical binary phase grating in the region M4 in three horizontal paths ($x$-direction). A state superposition of Gaussian modes paths, $\ket{1}_x\otimes\ket{0}_y$, $\ket{2}_x\otimes\ket{0}_y$ and $\ket{3}_x\otimes\ket{0}_y$, is considered to be our initial state. Each of the modes is reflected back in three different regions on the SLM: M1, M2 and M3. In the second incidence the beams may be diffracted by  linear phase gratings in which the first diffraction order propagates in three possible vertical directions, depending on the phase grating periodicity (mode paths: $\ket{i}_x\otimes\ket{1}_y$, $\ket{i}_x\otimes\ket{2}_y$ and $\ket{i}_x\otimes\ket{3}_y$, $i = 1, 2 \mbox{ and } 3$) . The multi-path system passes through a cylindrical lens which transforms all $\ket{i}_x\to\ket{2}_x$. After a spatial filtering, only the three possible first orders diffracted after the second incidence in the SLM reach the detection system. An  intensified charged-coupled device  (ICCD) camera  records the photon counts in each position.}
       \label{fig:Experimentalsetup}
\end{figure*}

The initial state $\ket{\psi_0}$ is prepared in three parallel photon paths, which are displaced relatively to each other in the $x$-direction and have a transverse Gaussian intensity profile. Fig. \ref{fig:Experimentalsetup} shows the experimental setup. The photon state preparation is realized by using a laser beam, attenuated to the single photon regime, that reaches a phase-only reflection SLM in the region M4 as shown in Fig. \ref{fig:Experimentalsetup}.
This region is programmed to have an periodical binary phase grating (BPG) in the $x$-direction. This gives rise to three Gaussian modes, selected by spatial filtering, displaced from each other in the $x$-direction \cite{PRABorges,PRABaldijao}. These three selected diffraction orders of higher intensity are labeled as: $\ket{1}_x\otimes\ket{0}_y$, $\ket{2}_x\otimes\ket{0}_y$ and $\ket{3}_x\otimes\ket{0}_y$. Our initial state can then be written as: $\ket{\psi_0}=\left(A_1e^{i\phi_{1}}\ket{1}_x+A_2e^{i\phi_{2}}\ket{2}_x+A_3e^{i\phi_{3}}\ket{3}_x\right)\otimes\ket{0}_y$, where the coefficients $A_i$ and the phases $\phi_i$, $i=1,2 \mbox{ and } 3$, depend only on the phase of BPG. These three paths are retro-reflected by a mirror to the same SLM and each path may be modulated by one of the three different periodical linear phase grating (LPG), in which the phase increases linearly in the $y$-direction. Each of these Gaussian beams, at the path $i$, that defines the initial state components $\ket{i}_x\otimes\ket{0}_y$ strikes a specified region $Mi$ ($i = 1, 2 \mbox{ and } 3$). We are able to choose from three different periodicities for those LPGs for producing three possible vertical ($y$-direction) displacements of the first diffracted order applied to each of the initial Gaussian modes $\ket{1}_x$, $\ket{2}_x$ and $\ket{3}_x$. In this way, we associate these three  beam positions in the $y$-direction with a different quantum state. We filter the first diffraction order only, which will be displaced vertically in relation to the initial state by either of the three established separation values. We label these path states as $\ket{1}_y$, $\ket{2}_y$ and $\ket{3}_y$. This operation transforms the state component $ \ket{i}_x\otimes\ket{0}_y \to \ket{i}_x\otimes\sum_{\ell=0}^{3}\beta_\ell\ket{\ell}_y$ ($i$ = 1, 2 and 3), where $\beta_\ell$ is a complex number which depends on the maximum phase and the periodicity of the LPG in the region $M_i$ of the SLM. A spatial filtering is applied for selecting the higher intensity first diffraction order in the $y$-direction. By using a cylindrical lens, we can merge the three paths in the $x$-direction into one path, the one labelled as the basis state $\ket{2}_x$. The final state $\ket{\psi}$ becomes:    
\begin{equation}
        \ket{\psi}=\ket{2}_x \otimes \sum^{3}_{\ell=1}\left(A_1\beta_{1\ell}e^{i\phi_{1l}}+A_2\beta_{2\ell}e^{i\phi_{2\ell}}+A_3\beta_{3\ell}e^{i\phi_{3\ell}}\right)\ket{\ell}_y
\end{equation}
Therefore, the ICCD camera detects only the light of the first orders diffracted in $y$-direction after the second incidence in the SLM. In our experimental apparatus we have as input state a photonic qutrit state prepared as three beams with transverse Gaussian profile displaced horizontally (path state in $x$-direction). The output qutrit state consists of three beams displaced vertically (path state in $y$-direction). By varying the diffraction gratings we can implement different operations \cite{PRABorges,PRABaldijao}. The large versatility of the realizable operations using this experimental scheme  allows us to simulate the three-level system decay dynamics. 
\subsection{\label{subsec:Imp_Kraus} Implementation of the Kraus operators} 

 We simulate the levels transitions in the three-level dynamics by changing the LPGs' periodicity among three chosen periods. The simulation of the transition to a $\ket{i}_y$ level ($i$ = 1, 2 \mbox{ and } 3) is achieved by selecting the first diffraction order of the three incident beams to LPGs' of different periodicities at the SLM.
 The transition probabilities among the three levels in the different decay configurations is modified by changing  the maximum phase of the LPGs in the $[0,2\pi ]$ phase interval, which controls the intensity of the diffracted light, \textit{i. e.}, the relative intensity among the diffraction orders. The larger the maximum phase the greater the amount of light in the first diffraction order. Different Kraus operators with a chosen $p_{ij}$ can be implemented by choosing different LPGs. All the Kraus operators were implemented by programming a frame sequence of different LPGs at the SLM, like a film (see Fig. \ref{fig:Experimentalsetup}). Each frame mimics a Kraus operator and its time duration was set in 100 ms. A complete SLM frame sequence lasts 300 ms for the $\lambda$ and $V$ decay dynamics and 400 ms for the cascade decay. The output state was recorded by an ICCD camera. The camera exposure time was set to be equal to the SLM frame sequence duration in each measurement.
 An average of 64 measurements was taken for a \{$p_{32}, p_{31}, p_{21}$\} set for characterizing the decay dynamics. 
By varying the $p_{ij}$-set we are able to simulate the time evolution for the dynamics of the three levels system decay since the $p_{ij}$ was parametrized by the evolution time $t$, as mentioned above. 

We assume the initial state to be pure and equal to $\ket{\psi_0}$ that can be rewritten as $\ket{\psi_0}=\frac{1}{\sqrt{I_T}}\left(\sqrt{I_1}\ket{1}+\sqrt{I_2}\ket{2}+\sqrt{I_3}\ket{3}\right)$, where $\frac{I_i}{I_T}= \vert A_i \vert ^2 $, $\phi_i$ = 0  and $I_T = \sum^{3}_{\ell=1} I_i$, with $i = 1, 2 \mbox{ and } 3$. Measurements at the image plane allow us to obtain the diagonal terms of the three-level density matrix, $\rho_{ii}$, which describes the relative population of level $\ket{i}$. For the complete decay dynamics characterization we need to measure the off-diagonal density matrix elements $\rho_{ij}$ that give information about the state coherence. By blocking one of the three beams we can measure two-beams interference patterns at the focal plane of a second cylindrical lens (see Fig. \ref{fig:Experimentalsetup}).
The two-beam interference patterns were fitted with Eq.~\ref{Eq:Int_pat} shown below. We obtain the off-diagonal terms of the density matrix from the visibilities of the two-beam patterns with $\mathcal{V}_{ij}=2\left|\sigma_{ij}(t)\right|$,where $\sigma_{ij}$ are the matrix elements of the new density operator in the Hilbert subspace $\{\ket{i},\ket{j}\}$ \cite{OPTMandel,PRAPaul,REFPAULA}. The experimental results are compared in the next subsection with the theoretical predictions.
\subsection{Detection}

The physical simulation of the atomic levels are implemented here by the photonic discrete Gaussian beam modes $\ket{l}_y$, represented in the position continuous space $\ket{l}_y=\int_{-\infty}^{\infty}dy exp\left[-\frac{\left(y-ld\right)^{2}}{2\sigma^2}\right]\ket{1y}$, where $l = 1, 2$ and $3$, $ld$ is the center position of each vertical spatial mode that are displaced by $d$ from its neighbour mode, $\sigma$ is the Gaussian mode transverse width, $\ket{1y} = \frac{1}{2\pi^2}\int_{-\infty}^{\infty}dq exp\left[-iqd\right]\ket{1q}$ is a representation of a photon Fock state in the transverse position and $\ket{1q}$ is the Fock state in the transverse momentum variable \cite{teich2001}. The probability of photon detection at the detection plane $z$ and transverse position $y$ is $P_y = Tr\left(\Gamma_y\rho\right)$,
with $\Gamma_y=E^{\left(-\right)}_y (y,z)E^{\left(+\right)}_y (y,z)$, where $\Gamma_y$ is the intensity operator which propagates the electromagnetic field from the SLM to the image plane,
$E^{\left(-\right)}_y (y,z)$ and $E^{\left(+\right)}_y (y,z)$ are the negative-frequency and
positive-frequency parts of the electric field operator at ($y,z$) \cite{mandel,leo07}, respectively. 
For the image plane measurements, the probability of photon detection in a position $y$ is
\begin{equation}
    Tr\left(\Gamma_y\rho\right)=\sum_{i=1}^{3}\rho_{ii}e^{-\frac{\left(y-id\right)^{2}}{\sigma^2}},
\end{equation}
where $\rho_{ii}$ are the diagonal terms of three-level system density matrix. The diagonal terms are the populations in each energy level and in the optical simulation $\rho_{ii}=\frac{I_i}{I_T}$, with $I_i$ being the intensity of the beam  $i$ ($i$ = 1, 2, 3) detected at the image plane by the ICCD and $I_T=I_1+I_2+I_3$. 

The decoherence effects are also characterized from the two-beam interference patterns obtained by blocking one of the beams and merging two of the three beams paths with a second cylindrical lens (Fig.~\ref{fig:Experimentalsetup}). Photons are then detected with the ICCD placed at the Focal plane of the second cylindrical lens. We define the matrix elements $\sigma_{ij}(t)$ which is obtained from the renormalized projection of $\rho(t)$ over the Hilbert subspace $\{\ket{i},\ket{j}\}$. Eq.~\ref{eq:reddensmatrix} shows the result of this operation on the density matrix for the cascade type dynamic $\rho_{c}(t)$ involving the $\ket{1}$ and $\ket{2}$ Hilbert subspace
    \begin{equation}
    \begin{aligned}
    \label{eq:reddensmatrix}
    \sigma(t)&=\frac{1}{I_r}
    \begin{pmatrix}
        I_1+I_2p+I_3p^2   &   \sqrt{I_1I_2}\sqrt{1-p}   &   0\\
        \sqrt{I_1I_2}\sqrt{1-p}   &   I_2(1-p)+I_3p(1-p)   &  0\\
        0 & 0 &  0
    \end{pmatrix}\\ \\
    \end{aligned},
    \end{equation}
where $I_r=I_1+I_2+pI_3$ and  we considered that $p=p_{ij}~\forall~i,j$.
The probability of detection in the Fourier plane proportional to the expected value of $\Gamma^{\prime}_y$, which is the Fourier transform of the intensity operator, $\Gamma^{\prime}_y$, and propagates the electromagnetic field from the SLM to the Fourier plane. Thus we have the interference pattern of the state in Eq.~\ref{eq:reddensmatrix} which is given by
 \begin{equation}\label{Eq:Int_pat}
    Tr\left(\Gamma_y^{\prime}\sigma\right)=e^{-\frac{-\sigma k^2 y^{2}}{f^2}}\left[1+2\left|\sigma_{ij}\right|cos\left(\frac{kyd}{f}+\phi_{ij}\right)\right],
\end{equation}
where $f$ is the focal length of the lens, $k$ is the modulus of the wave vector and $\sigma_{ij}=\left|\sigma_{ij}\right|e^{i\phi_{ij}}$. Thus the visibility of the interference pattern between the paths $\ket{i}$ and $\ket{j}$ becomes $\mathcal{V}_{ij}=2\left|\sigma_{ij}(t)\right|$ \cite{OPTMandel,PRAPaul}.
\section{\label{sec:level3}Experimental results and discussion}
\begin{figure}[htb!]     \includegraphics[scale=0.33]{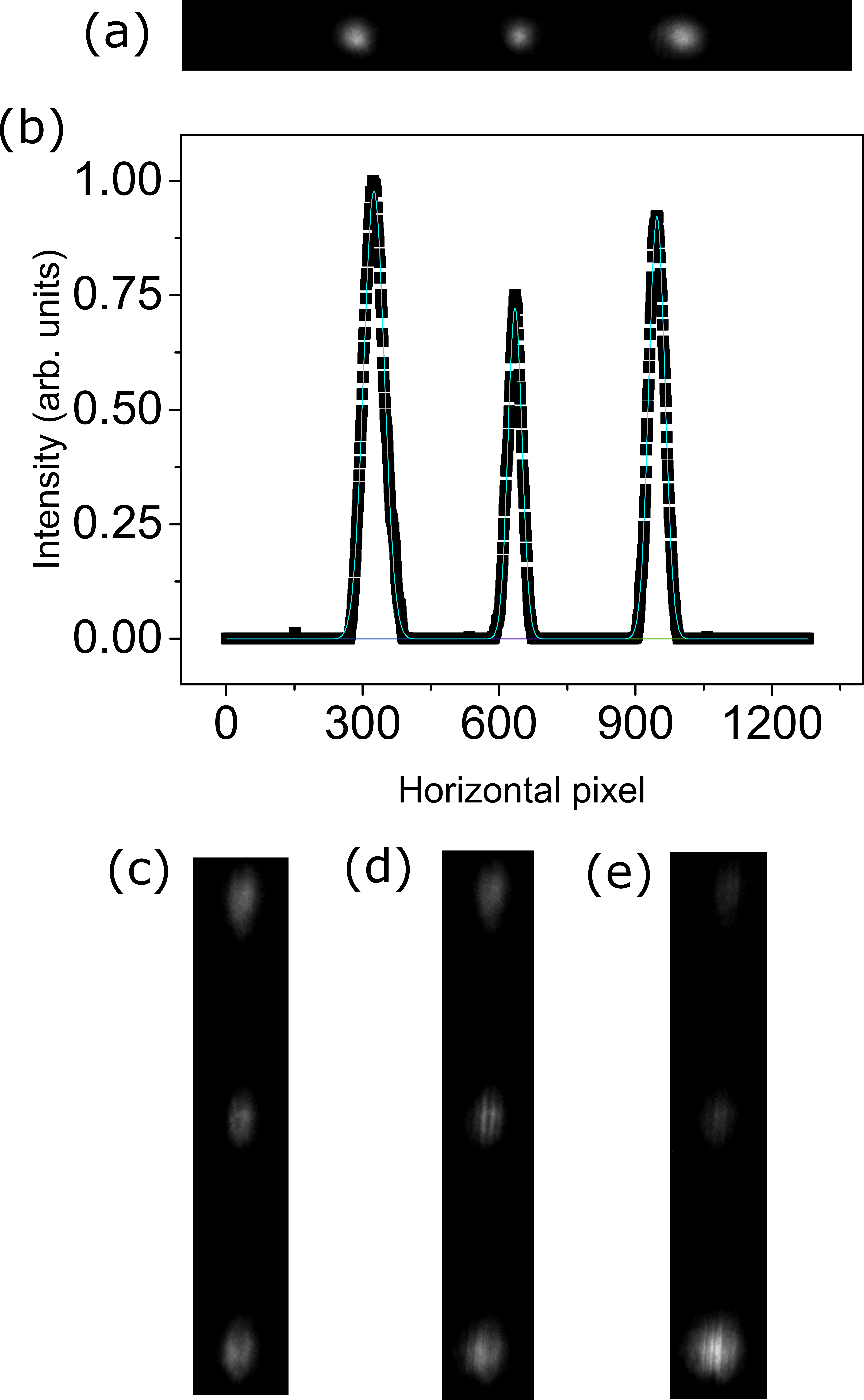}
        \caption{(a) Image measurement of the initial state for the cascade dynamics. (b) Sum over the photon counts in the columns for obtaining a normalized vertically integrated optical profile (ITOP) of the initial state for the cascade dynamics. And image measurements of the implementation of the cascade dynamics for: (c) $p=0$; (d) $p=0.25$ and (e) $p=0.75$}.
    \label{Imageandgraphinitialstate-cascade}
\end{figure}

\begin{figure}[htb!]
      \includegraphics[scale=0.39]{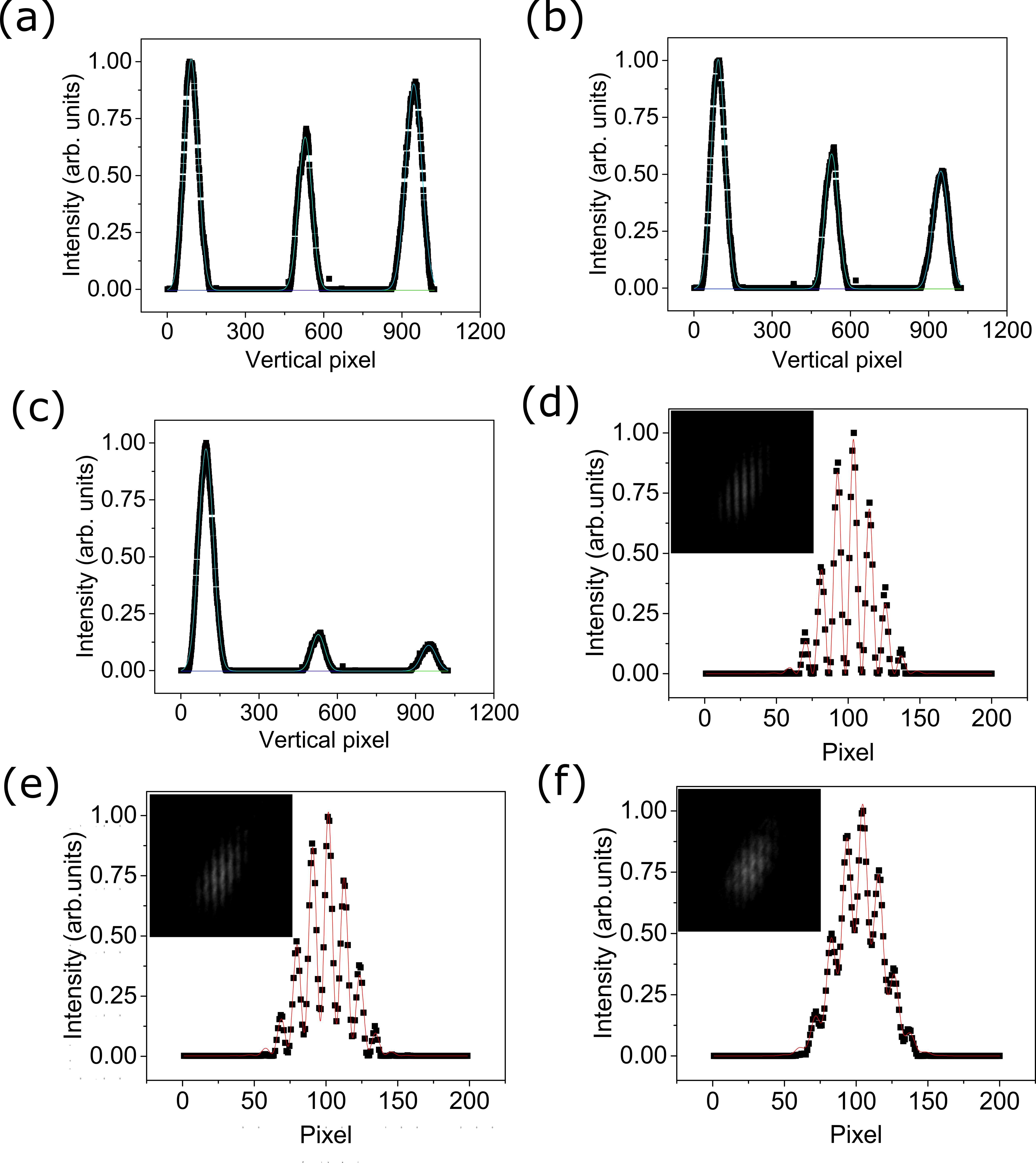}
        \caption{Normalized vertically integrated optical profle (ITOP) of the image measurements of the cascade dynamics for: (a) $p=0$; (b) $p=0.25$ and (c) $p=0.75$. Interference patterns between the path states $\ket{1}$ and $\ket{2}$, with its ITOPs of the cascade dynamics for: (d) $p=0$; (e) $p=0.25$ and (f) $p=0.75$ . The incidence plane is rotated in our setup which demands that all the patterns must be rotated clockwise in the data analysis.}
    \label{fig:Interferencemeasurements-cascade}
\end{figure}

The initial state prepared for the cascade dynamics simulation is generated by a BPG. The three beams image detected by the ICCD is shown in the Fig. \ref{Imageandgraphinitialstate-cascade} (a). The three photonic beams represent the three-level system and the photon count in each beam represent the level population. By means of a MATLAB program, we make a sum over the photon counts in the columns and obtain a normalized vertically integrated transverse optical profile (ITOP), plotted in Fig. \ref{Imageandgraphinitialstate-cascade} (b) in terms of the horizontal pixels.
After an image characterization of our initial state, we  proceed to the implementation of the cascade dynamics. By preparing all the diffraction gratings that simulate the Kraus operators in a film at the SLM, we measure the image of the three vertical beams (path Gaussian states) as shown in Fig. \ref{fig:Experimentalsetup}. Fig. \ref{Imageandgraphinitialstate-cascade} illustrates the image obtained with the implementation of the Kraus operators that describe the cascade dynamics for some values of $p$ between 0 and 1 with incremental step of $0.125$.
For a complete coherence characterization of the photon path three-level state that occurs during the process of decay, we also measure the visibility of the interference patterns of the photon beam pairs (the beams were measured two by two), for the same probability values. This was realized by inserting an extra cylindrical lens (Fig. \ref{fig:Experimentalsetup}) that focus the beams in the vertical direction at the ICCD plane. The interference patterns between the beams that represent the path states $\ket{1}$ and $\ket{2}$ in the cascade dynamics are shown in Fig. \ref{fig:Interferencemeasurements-cascade}.

\begin{figure}[t!]
         \includegraphics[scale=0.37]{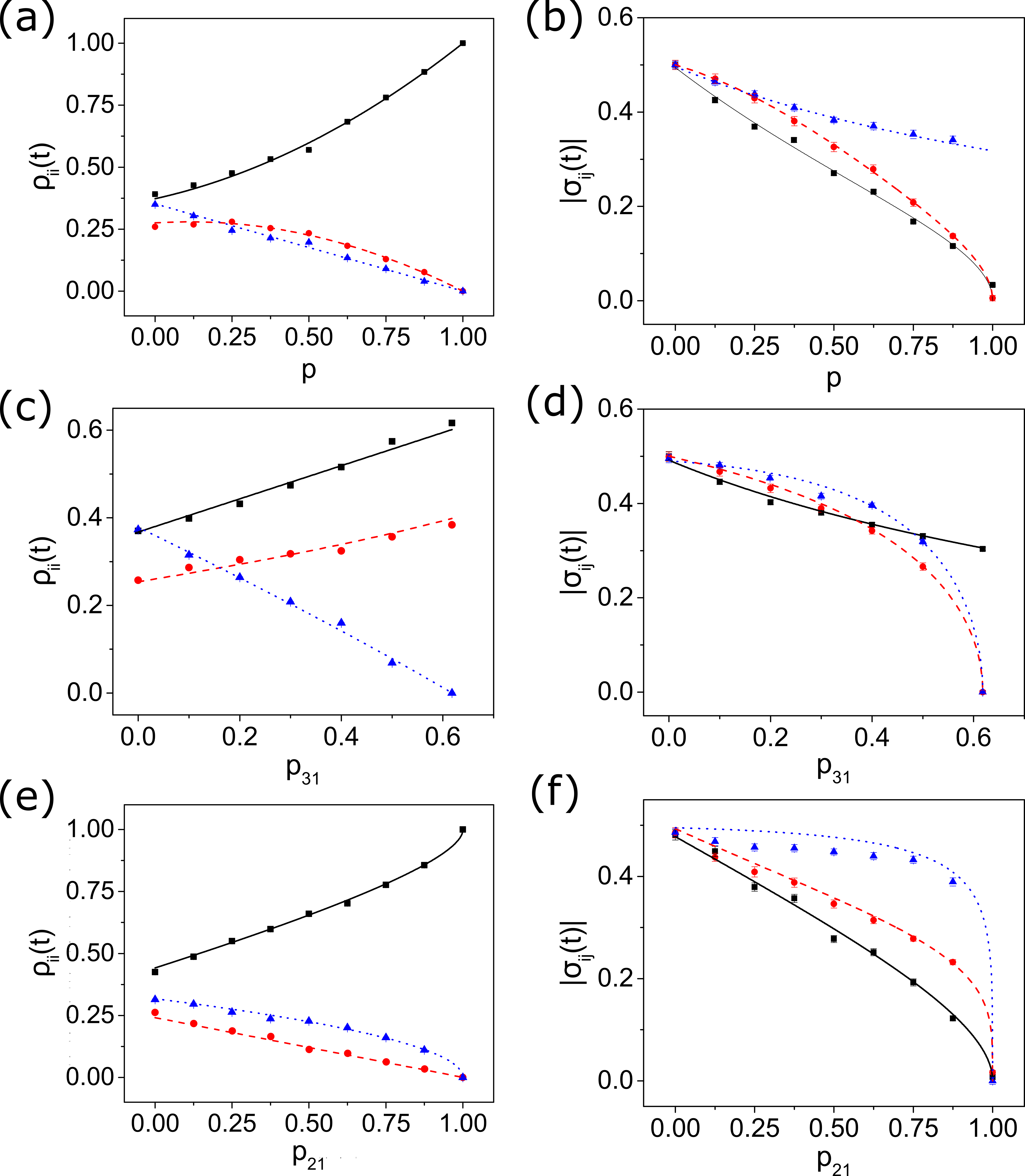}
        \caption{Time evolution of the density matrix elements: (a) diagonal elements of the density matrix for the cascade dynamics; (b) modulus of the off-diagonal elements of the density matrix for the cascade dynamics; (c) diagonal elements of the density matrix for the $\Lambda$ dynamics; (d)  modulus of the off-diagonal elements of the density matrix for the $\Lambda$ dynamics; (e) diagonal elements of the density matrix for the $V$ dynamics; and (f)  modulus of the off-diagonal elements of the density matrix for the $V$ dynamics. In the diagonal (off-diagonal) graphs the lines correspond to the theoretical predictions and the symbols are the measurement values, where black is $\rho_{11}(t)$ ($\left|\sigma_{12}(t)\right|$), red is $\rho_{22}(t)$ ($\left|\sigma_{13}(t)\right|$) and blue is $\rho_{33}(t)$ ($\left|\sigma_{23}(t)\right|$).}
    \label{fig:Timeevolutionfordensitymatrixelements}
\end{figure}

 The measurements at the image planes give us the diagonal elements of the matrix density (area below the ITOPs curves). By performing the measurements at the image plane and at the focal plane (interference patterns), we are able to obtain the off-diagonal elements of the three-level density matrix. From the visibility of the two-beam interference patterns, we obtain the modulus of the off-diagonal elements of the matrix density. Those results are shown in the Fig. \ref{fig:Timeevolutionfordensitymatrixelements} where the modulus of the density operator elements are plotted in terms of $p$ (function of the decay time $t$). The continuous curve is the theoretical prediction obtained by using the Kraus operators for deriving the density operator evolution.

For the $\Lambda$ dynamics, we proceeded  analogously to the cascade dynamics to simulate and analyse the three-level decay. We made $\gamma_{31}=2\gamma_{32}$, aiming to have different values of $p_{31}$ and $p_{32}$ for $t\neq0$. We characterized the $\Lambda$ decay from $p_{31} = p_{32}= 0 $ to the end of the decay dynamics when no population exists in level 3, \textit{i.e.}, when $p_{31}= 1 - p_{32}$. The results for the time evolution of the diagonal terms and the modulus of the off-diagonal ones of the density matrix is depicted in Fig.
\ref{fig:Timeevolutionfordensitymatrixelements}.

Finally, for the $V$ dynamics, we followed the same steps for the other two previous decay dynamics and we considered $\gamma_{21}=2\gamma_{31}$. The results for the time evolution of the diagonal terms and the modulus of the off-diagonal ones of the three-level density matrix is depicted in Fig.~\ref{fig:Timeevolutionfordensitymatrixelements}.
 In spite of some small deviations of the theoretical prediction, the theoretical continuous lines fit the measured points very well and we were able to simulate the three-level dynamics satisfactorily. 
\section{\label{conclusions}Conclusion}
We simulate the three-level state dynamics by using photonic qutrit path state for representing the three-level system. The qutrit state is prepared initially by generating a photon superposition state of three-path Gaussian modes. Programming a temporal sequence of phase spatial gratings, a SLM diffracts the Gaussian modes and implements Kraus operators that describe the three-level decay dynamics. Therefore, the SLM is responsible for implementing generalized quantum operations on the initial three-level photonic quantum state.
The experimental characterizations of the density matrix for the output state are in agreement to the theoretical predictions despite a small deviation. With a precise periodical phase modulation of the photonic paths, we are able to implement a large number of operations and simulate the different dynamics of decay in a three-level system. This simulation give us a better comprehension of how quantum jump affects the coherence of a three-level system. Moreover, this implementation can be used for understanding how quantum jumps in high dimension systems affects quantum protocols due to the state decoherence. This procedure is general and it can be extended to multilevel systems with a number of levels higher than three with pixel number and beam shape restrictions.

The quantity of achievable quantum operations with this apparatus makes it a good choice for other implementations. For instance, we are able to simulate an atomic system being excited by an external electromagnetic field, not only the decaying. We can also test the behavior of a quantum channel subjected to this kind of noise.
\section*{ACKNOWLEDGMENTS}

We are grateful to Conselho Nacional de Desenvolvimento Cient\'\i{}fico e Tecnol\'ogico (CNPq), Instituto Nacional de Ci\^encia e Tecnologia de Informa\c{c}\~ao Qu\^antica (INCT-IQ), Coordena\c{c}\~ao de Aperfei\c{c}oamento de Pessoal de N\'ivel Superior (CAPES), and Funda\c{c}\~ao de Amparo \`a Pesquisa do Estado de Minas Gerais (FAPEMIG). S.P. acknowledges the CAPES-PRINT for supporting his stay as visiting professor at University of Glasgow during the preparation of this work.



\begin{thebibliography}{0}%
\makeatletter
\providecommand \@ifxundefined [1]{%
 \@ifx{#1\undefined}
}%
\providecommand \@ifnum [1]{%
 \ifnum #1\expandafter \@firstoftwo
 \else \expandafter \@secondoftwo
 \fi
}%
\providecommand \@ifx [1]{%
 \ifx #1\expandafter \@firstoftwo
 \else \expandafter \@secondoftwo
 \fi
}%
\providecommand \natexlab [1]{#1}%
\providecommand \enquote  [1]{``#1''}%
\providecommand \bibnamefont  [1]{#1}%
\providecommand \bibfnamefont [1]{#1}%
\providecommand \citenamefont [1]{#1}%
\providecommand \href@noop [0]{\@secondoftwo}%
\providecommand \href [0]{\begingroup \@sanitize@url \@href}%
\providecommand \@href[1]{\@@startlink{#1}\@@href}%
\providecommand \@@href[1]{\endgroup#1\@@endlink}%
\providecommand \@sanitize@url [0]{\catcode `\\12\catcode `\$12\catcode
  `\&12\catcode `\#12\catcode `\^12\catcode `\_12\catcode `\%12\relax}%
\providecommand \@@startlink[1]{}%
\providecommand \@@endlink[0]{}%
\providecommand \url  [0]{\begingroup\@sanitize@url \@url }%
\providecommand \@url [1]{\endgroup\@href {#1}{\urlprefix }}%
\providecommand \urlprefix  [0]{URL }%
\providecommand \Eprint [0]{\href }%
\providecommand \doibase [0]{http://dx.doi.org/}%
\providecommand \selectlanguage [0]{\@gobble}%
\providecommand \bibinfo  [0]{\@secondoftwo}%
\providecommand \bibfield  [0]{\@secondoftwo}%
\providecommand \translation [1]{[#1]}%
\providecommand \BibitemOpen [0]{}%
\providecommand \bibitemStop [0]{}%
\providecommand \bibitemNoStop [0]{.\EOS\space}%
\providecommand \EOS [0]{\spacefactor3000\relax}%
\providecommand \BibitemShut  [1]{\csname bibitem#1\endcsname}%
\let\auto@bib@innerbib\@empty
\end{thebibliography}%


\begin{thebibliography}{99}
 
 \bibitem{flamini} Fulvio Flamini, Nicolas Spagnolo, and Fabio Sciarrino, Rep. Prog. Phys. \textbf{82} 016001 (2019).
 \bibitem{slussarenko} Sergei Slussarenko and  Geoff J. Prydeb, Appl. Phys. Rev. \textbf{6}, 041303 (2019).
\bibitem{hensen} B. Hensen, et. al., Nature  \textbf{526}, 682 (2015); M. Giustina, et. al., Phy. Rev. Lett.  \textbf{115}, 250401 (2015); Lynden K. Shalm, et. al., Phys. Rev. Lett.  \textbf{115}, 250402 (2015). 
\bibitem{ralph} P. Kok, W. J. Munro, K. Nemoto, T. C. Ralph, J. P. Dowling, and G. J. Milburn, Rev. Mod. Phys. \textbf{79}, 135 (2007), T. Ralph and G. Pryde,  Prog. Opt. \textbf{54}, 209 (2010).

\bibitem{petruccione}H.-P. Breuer and F. Petruccione, \textit{The Theory of Open Quantum Systems} (Oxford University Press, Oxford, 2002).
\bibitem{alicki} R. Alicki and K. Lendi, \textit{Quantum Dynamical Semigroups and Applications} (Springer Science $\&$ Business Media, 2007).
\bibitem{carmichael} H. Carmichael, \textit{ An Open Systems Approach to Quantum Optics}, Springer-Verlag, Berlin (1993).
\bibitem{REVZurek} W. H. Zurek, Reviews of Modern Physics, \textbf{75}, 715 (2003); W. H. Zurek, 
Phys. Rev. D \textbf{26} 1862 (1982); J. P. Paz, W. H. Zurek, \textit{Environment-induced decoherence and the transition from quantum to classical}, in: R. Kaiser, C. Westbrook, F. David (Eds.), Coherent Atomic Matter Waves, Les Houches Session LXXII, Vol. 72 of Les Houches Summer School Series, Springer, Berlin, 533 (2001).
\bibitem{marquardt} F. Marquardt and A. Puttmann, \textit{Introduction to dissipation and decoherence in quantum systems}, arxiv.org:0809.4403 (2008).
\bibitem{sc}M. Scholosshauer, Phys. Rep. \textbf{831}, 1 (2019).
\bibitem{SCIFarias}O. Jim{\'e}nez Far{\'\i}as, C. Lombard Latune, S. P. Walborn, L. Davidovich and P. H. Souto Ribeiro; Science \textbf{324}, 5933 (2009).
\bibitem{SCIREPMarques}B. Marques, A. A. Matoso, W. M. Pimenta, A. J. Gutiérrez-Esparza, M. F. Santos and S. P{\'a}dua; Sci. Rep. \textbf{5}, 16049 (2015).
\bibitem{SCIENCEAlmeida}M. P. Almeida, F. De Melo, M. Hor-Meyll, A. Salles, S. P. Walborn, P. H. Souto Ribeiro and L. Davidovich; Science \textbf{316}, 5824 (2007).
\bibitem{unden}T. Unden, et. al.; Phys. Rev. Lett. \textbf{116}, 230502 (2016). 
\bibitem{mataloni} A. Chiuri, V. Rosati, G. Vallone, S. P{\'a}dua, H. Imai, S. Giacomini, C. Macchiavello, and P. Mataloni; Phys. Rev. Lett. \textbf{107}, 253602 (2011).
\bibitem{aolita} L. Aolita, F. De Melo, L. Davidovich; Rep. on Prog. in Phys. \textbf{78} 042001 (2015).
\bibitem{almeida} A. Salles, F. de Melo, M. P. Almeida, M. Hor-Meyll, S. P. Walborn, P. H. Souto Ribeiro, and L. Davidovich;
Phys. Rev. A \textbf{78} 022322 (2008).
\bibitem{laurat}J. Laurat, K. S. Choi, H. Deng, C. W. Chou, and H. J. Kimble;  Phys. Rev. Lett. \textbf{99} 180504 (2007).
\bibitem{barbosa}F. A. S. Barbosa, A. S. Coelho, A. J. de Faria, K. N. Cassemiro, A. S. Villar, P. Nussenzveig, and M. Martinelli, 
Nat. Photon., \textbf{4} 858  (2010).
\bibitem{xu}J-S. Xu, C-F. Li, X-Y. Xu, C-H. Shi, X-B. Zou, and G-C. Guo, 
Phys. Rev. Lett., \textbf{103} 240502 (2009).
\bibitem{blatt} H. H\"affner, F. Schmidt-Kaler, W. H \"ansel, C. F. Roos, T. K\"orber, M. Chwalla, M. Riebe, J. Benhelm, U. D. Rapol, C. Becher, and R. Blatt,
Appl. Phys. B \textbf{81}, 151 (2005).
\bibitem{cook}Richard J. Cook, Phys. Scr. \textbf{1988} 49 {1988};  J. Dalibard, Y. Castin, K. Mølmer, Phys. Rev. Lett. \textbf{68} 580 (1992); Nicolas Gisin and Ian C. Persival, Phys. Lett. A \textbf{167}, 315 (1992);N. Gisin, P. L. Knight, I. C. Percival, R. C. Thompson, and D. C. Wilson, J. of Mod. Opt-, \textbf{40}  1663 (1993);
M. B. Plenio, P. L. Knight, Rev. of Mod. Phys. \textbf{70} 101 (1998).
\bibitem{dick} R. Dick, Studies in History and Philosophy of Modern Physics \textbf{57}115 (2017).
\bibitem{bohr} N. Bohr (1913), Phil. Magazine \textbf{26}, 476 (1993).
\bibitem{sauter} T. H. Sauter, W.  Neuhaser, R. Blatt, and P. E. Toscheck, Phys. Rev. Lett. \textbf{57}, 1696 (1986); J. C. Bergquist, R. G. Hulet, W. M. Itano, and D. J. Wineland, Phys . Rev . Lett . \textbf{57}, 1699 (1986),
W. Nagourney, J. Sandberg, and H. Dehmelt, Phys . Rev . Lett . \textbf{56}, 2797 (1986); Randall G. Hulet, D. J. Wineland, J. C. Bergquist, and Wayne M. Itano, Phys. Rev. A \textbf{37}, 4544(R) (1988).
\bibitem{basch} T. Basch\'e, S. Kummer, and C. Brauchle,  Nature \textbf{373}, 132 (1995). 
\bibitem{peil}S. Peil and G. Gabrielse
Phys. Rev. Lett. \textbf{83}, 1287 (1999).
\bibitem{gleyzes}S. S. Gleyzes, S. Kuhr, C. Guerlin, J. Bernu, S. Deléglise, U. Busk, M. Brune, J.-M. Raimond, S. Deleglise, U. Busk Hoff, M. Brune, J.-M. Raimond, and S. Haroche Nature \textbf{446}, 297 (2007).
\bibitem{solid}F. Jelezko, I. Popa, A. Gruber, C. Tietz, and J. Wrachtrup, Appl. Phys. Lett. \textbf{81}, 2160 (2002); P. Neumann, J. Beck, M. Steiner, F. Rempp, H. Fedder, P. R. Hemmer, J. Wrachtrup, F. Jelezko, Science \textbf{329}, 542 (2010); L. Robledo, L. Childress, H. Bernien, B. Hensen, P. F. A. Alkemade, and R. Hanson, Nature \textbf{477} 574 (2011); R. Vijay, D. H. Slichter, and I. Siddiqi (2011), Phys. Rev. Lett. \textbf{106} 110502 (2011); M. Hatridge, S. Shankar, M. Mirrahimi, F. Schackert, K. Geerlings, T. Brecht, K. M. Sliwa, B. Abdo, L. Frunzio, S. M. Girvin, R. J. Schoelkopf, and M. H. Devoret, Science \textbf{339} 178 (2013).
\bibitem{photo} M. Ossiander, J. Riemensberger, S. Neppl, M. Mittermair, M. Schäffer, A. Duensing, M. S. Wagner, R. Heider, M. Wurzer, M. Gerl, M. Schnitzenbaumer, J. V. Barth, F. Libisch, C. Lemell, J. Burgdörfer, P. Feulner an R. Kienberger, Nature \textbf{561}, 374 (2018).
\bibitem{lasercooling}Paul D. Lett, Richard N. Watts, Christoph I. Westbrook, and William D. Phillips, Phillip L. Gould and Harold J. Metcalf, Phys. Rev. Lett. \textbf{11}, 169 (1988);A. Aspect, E. Arimondo, R. Kaiser, N. Vansteenkiste, and C. Cohen-Tannoudji, Phys. Rev. Lett. \textbf{61}, 826 (1988); J. Dalibard and C. Cohen-Tannoudji, JOSA B \textbf{6}, 2023 (1989);Mark Kasevich and Steven Chu, Phys. Rev. Lett. \textbf{69}, 1741 (1992);R. Gupta, C. Xie, S. Padua, H. Batelaan, and H. Metcalf, Phys. Rev. Lett. \textbf{71} 3087 (1993).
\bibitem{enhance}P. Goy, J. M. Raimond, M. Gross, and S. Haroche, Phys. Rev. Lett. \textbf{50}, 1903 (1983); W. Jhe, A. Anderson, E. A. Hinds, D. Meschede, L. Moi, and S. Haroche, Phys. Rev. Lett. \textbf{58}, 666 (1987); F. De Martini, G. Innocenti, G. R. Jacobovitz, and P. Mataloni, Phys. Rev. Lett. \textbf{59}, 2955 (1987).
\bibitem{minev}Z. K. Minev, S. O. Mundhada, S. Shankar, P. Reinhold, R. Gutiérrez-Jáuregui, R. J. Schoelkopf, M. Mirrahimi, H. J. Carmichael and M. H. Devoret, Nature \textbf{570}, 200 (2019); Z. K. Minev, \textit{Catching and Reversing a Quantum Jump Mid-Flight}, Ph. D Thesis, Yale University (2018) (arxiv:1902.10355 (2019)).
\bibitem{sonja}J. Leach, M. J. Padgett, S. M. Barnett, S. Franke-Arnold, J. Courtial
Phys. Rev. Lett. \textbf{88}, 25790 (2002).
\bibitem{padgett} A. C. Dada, J. Leach, G. S.Buller, M. J. Padgett, E. Andersson, Nat. Phys. \textbf{7}, 677 (2011).
\bibitem{langford} N. K. Langford, R. B. Dalton, M. D. Harvey, J. L. O’Brien, G. J. Pryde, S. D. Bartlett, and A. G. White, Phys. Rev. lett. \textbf{93}, 053601 (2002).
\bibitem{zeilinger}A.  Mair,  A.  Vaziri,  G.  Weihs,  A.  Zeilinger,  Nature \textbf{412},  313 (2001).
\bibitem{PRLNeves} L. Neves, G. Lima, and J. G. Aguirre G\'omez, C. H. Monken, C. Saavedra, and S. P\'adua, Phys. Rev. Lett. \textbf{94}, 100501 (2005).
\bibitem{lima09} G. Lima, A. Vargas, L. Neves, R. Guzm\'an, and C. Saavedra, Opt. Express \textbf{17}, 10688 (2009).
\bibitem{stevereview} S. P. Walborn, C. H. Monken, S. P\'adua, and P. H. Souto Ribeiro, Physics Reports \textbf{495} 87 (2010).
\bibitem{PRABorges} G. F. Borges, R. D. Baldij\~ao, and J. G. L. Cond\'e, J. S. Cabral, B. Marques, M. Terra Cunha, A. Cabello and S. P\'adua; Phys. Rev. A \textbf{97}, 022301 (2018).
\bibitem{guo}Xiao-Min Hu, Jiang-Shan Chen, Bi-Heng Liu, Yu Guo, Yun-Feng Huang, Zong-Quan Zhou, Yong-Jian Han, Chuan-Feng Li, and Guang-Can Guo, Phys. Rev. Lett. \textbf{117}, 170403 (2016).
\bibitem{leo1} M. A. Sol\'is-Prosser, A.  Arias, J. J. M. Varga, L. Rebón, S. Ledesma, C. Iemmi, and L. Neves, Optics Letters,  \textbf{38}, 4762 (2013).
\bibitem{holanda} W. H. Peeters, J. J. Renema and M. P. van Exter, Phys. Rev. A \textbf{79}, 043817 (2009).
\bibitem{paula} P. Machado, A. A. Matoso, M. R. Barros, L. Neves, and S P\'adua, Phys. Rev. A 99 \textbf{6}, 063839 (2019).
\bibitem{pierre} P.-L. de Assis, M. A. D. Carvalho, L. P. Berruezo, J. Ferraz, I. F. Santos, F. Sciarrino and S. Pádua, Opt. Express \textbf{19} 3715 (2011).
\bibitem{PRAMarques} B. Marques, M. R. Barros, W. M. Pimenta, M. A. D. Carvalho, J. Ferraz, R. C. Drumond, M. Terra Cunha and S. P\'adua, Phys. Rev. A \textbf{86}, 032306 (2012).
\bibitem{miguelleo} M. A. Sol\'is-Prosser,  M. F. Fernandes, O. Jim\'enez, O. Delgado, and L. Neves, Phys. Rev. Lett., \textbf{118}, 100501 (2017). 
\bibitem{boyd} H. Larocque, J. Gagnon-Bischoff, D. Mortimer, Y. Zhang, F. Bouchard,J. Upham, V. Grillo, R. W. Boyd, and E. Karimi,
Optics express \textbf{25} 19832 (2017).
\bibitem{nori} I. M. Georgescu, S. Ashhab, and F. Nori, Rev. Mod. Phys. \textbf{86} 153 (2014).
 \bibitem{blatt2}J. T. Barreiro, M. Muller, P. Schindler, D. Nigg, T. Monz, M. Chwalla, M. Hennrich, Christian F. Roos, P. Zoller, and R. Blatt, Nature \textbf{470} 486 (2011).
\bibitem{PRLLukin}M. D. Lukin, M. Fleischhauer, R. Cote; Phys. Rev. Lett. \textbf{87}, 037901 (2001).
\bibitem{PRABaldijao}R. D. Baldij\~ao, G. F. Borges, B. Marques, M. A. Sol\'{\i}s-Prosser, L. Neves, S. P\'adua; Phys. Rev. A \textbf{96}, 032329 (2017).
\bibitem{OPTMandel}L. Mandel; Opt. Lett. \textbf{16}, 1882 (1991).
\bibitem{PRAPaul}T. Paul, T. Qureshi; Phys. Rev. A \textbf{95}, 042110 (2017).
\bibitem{REFPAULA}P. Machado and S. P\'adua; arXiv:1910.09574v1.
\bibitem{teich2001} A. F. Abouraddy, B. E. A. Saleh, A. V. Sergienko, and M. C.
Teich, Phys. Rev. Lett. \textbf{87}, 123602 (2001).
\bibitem{mandel} Leonard Mandel and Emil Wolf \textit{Optical Coherence and Quantum Optics} Cambridge University Press (2013)
\bibitem{leo07}L. Neves, G. Lima, E. J. S. Fonseca, L. Davidovich, and S. Pádua, Phys. Rev. A \textbf{76}, 032314 (2007).

\end{thebibliography}
\end{document}